\documentclass[12pt,draftclsnofoot, onecolumn]{IEEEtran}
\usepackage{graphicx,cite,epsfig,amssymb,amsmath,subfigure,url,stfloats,latexsym}
\usepackage{multirow,array}
\usepackage{graphicx,cite,amsmath,amssymb,hhline,booktabs,stfloats,bm}
\usepackage{verbatim}
\usepackage[utf8]{inputenc}
\usepackage{subfigure}
\usepackage{algorithm,algorithmic}
\usepackage{url}
\usepackage{xcolor}
\usepackage{epstopdf}

\begin{document}
	\title{Latency Minimization for Task Offloading in Hierarchical Fog-Computing C-RAN Networks}
	\author{\IEEEauthorblockN{Yijin Pan\IEEEauthorrefmark{1}\IEEEauthorrefmark{2}, 
			Huilin Jiang\IEEEauthorrefmark{3},
			Huiling Zhu\IEEEauthorrefmark{2},
			and Jiangzhou Wang\IEEEauthorrefmark{2}}
		\IEEEauthorblockA{\IEEEauthorrefmark{1}National Mobile Communications Research Laboratory, Southeast University, China}
		\IEEEauthorblockA{\IEEEauthorrefmark{2}School of Engineering and Digital Arts, University of Kent, UK.}
		\IEEEauthorblockA{\IEEEauthorrefmark{3}School of Electronic Engineering, Nanjing Xiaozhuang University, Nanjing, China.} \\
		Email: panyj@seu.edu.cn, huilin.jiang@njxzc.edu.cn,\{y.pan,h.zhu,j.z.wang\}@kent.ac.uk}
	\maketitle
	
	\begin{abstract}
		Fog-computing network combines the cloud computing and fog access points (FAPs) equipped with mobile edge computing (MEC) servers together to support computation-intensive tasks for mobile users.
		However, as FAPs have limited computational capabilities and are solely assisted by a remote cloud center in the baseband processing unit (BBU) of the cloud radio access (C-RAN) network, the latency benefits of this fog-computing C-RAN network may be worn off when facing a large number of offloading requests. 
		In this paper, we investigate the delay minimization problem for task offloading in a hierarchical fog-computing C-RAN network, which consists of three tiers of computational services: MEC server in radio units, MEC server in distributed units, and the cloud computing in central units.
		The receive beamforming vectors, task allocation, computing speed for offloaded tasks in each server and the transmission bandwidth split of fronthaul links are optimized by solving the formulated mixed integer programming problem. 
		The simulation results validate the superiority of the proposed hierarchical fog-computing C-RAN network in terms of the delay performance.
	\end{abstract}

	
	\IEEEpeerreviewmaketitle

	\section{Introduction}
	
	The cloud radio access network (C-RAN) architectures have been suggested as the prospective solution for the fifth-generation (5G) and Beyond (B5G) networks\cite{wu2015cloud}. 
	The C-RAN architecture completely breaks down the traditional structures of base stations, which are separated into the radio units (RUs), a baseband processing unit (BBU) or central unites (CUs)\cite{zhu2011performance}, and the optical or wireless connection fronthaul link between them\cite{7880689,7938594,wang2012distributed,zhu2013radio,wei2015mutual}.
	This configurable structure provides reduced operational costs and ubiquitous access to the shared pools of processing capabilities and storage resources.
	
	Nevertheless, when handling the emerging computation-intensive and delay-sensitive applications, such as the virtual reality and automatic driving, the cloud computing approach integrated in BBU cannot provide satisfactory end-to-end delay performance due to the far-away transmission distances\cite{Wang.2018}. 
	Hence, a new paradigm is being considered by pushing some functions of the BBU and computational resources to the network’s edge in the users' vicinity, which is called edge computing, or the fog-computing paradigm\cite{8100873}.

	The fog-computing paradigm is normally regarded as a cloud-assisted  mobile edge computing (MEC) system, which consists of two tiers: fog access points (FAPs) such as the RUs with storage and processing functionalities, and the cloud center connected to BBU\cite{Rahman.2018,Liu.2018}. 
	This fog-computing paradigm has been well studied in the literature in regards to improvements in task offloading performance\cite{Zhao.2019,Du.2018}.
	For instance, the latency minimization strategy was investigated in \cite{Rahman.2018} by forming FAP grouping and considering the caching function.
	In \cite{Liu.2018}, the tasks scheduling strategy and the resource allocation scheme was investigated in fog-computing network with non-orthogonal multiple access. 
	A min-max fairness based cost minimization problem was addressed in \cite{Du.2018} by considering a mixed fog/cloud computing system via a joint optimization of offloading decision making and resource allocation.
	It was shown in \cite{AbdelAtty.2019} that improving the delay performance by using a fog architecture is not a straightforward process but rather requires particular care in terms of choosing the appropriate mode when placing/installing fog functions in fog devices.
	
	However, in the aforementioned fog-computing paradigm,
	the edge FAPs suffer from a limited computational capability that may offset the latency benefits, and thus cannot cope with a vast amount of offloading requests\cite{AbdelAtty.2019}. 
	Moreover, a new splitting partition of BBU functions between the CU and DU (distributed units) has been suggested for the evolved C-RAN network for the sake of high-rate transmission \cite{Gomes.2018}, as the high speed transport of sampled radio waveforms between the BBU and the RUs will no longer be feasible \cite{3GPP.2017}. 
	This inspires us to consider a hierarchical fog-computing system, where the MEC servers with different computation capabilities are provisioned in DUs and RUs of the C-RAN network. 
	This multi-tier fog-computing scheme has the potential to improve the latency performance due to the reduced transmission hops and distances of task offloading, in contrast to the aforementioned fog-computing diagrams that solely rely on the assistance of a remote cloud center in BBUs.
	Although a similar hierarchical fog-computing structure have been considered very recently in \cite{ElHaber.2019} and \cite{Wu.2019}, the benefits of this structure has not been fully evaluated, especially in terms of the delay performance.
	In \cite{ElHaber.2019}, joint power control and resources allocation was investigated to minimize the computational cost and energy consumption, but the improvement of delay performance achieved by this structure was not elaborated.
	In the offloading approach proposed in \cite{Wu.2019}, the computation disparity between tiers and the transmission scheduling of the bandwidth limited fronthaul was not exploited and optimized, which may lead to inefficient solutions.

	In fact, revealing how the system parameters affect the delay performance, such as the different computation capabilities in different tiers and the bandwidth of the fronthaul links, can provide important guidance for the network operators to reduce unnecessary costs without affecting network performance.
	Therefore, it is of practical significance to investigate the task scheduling and computational/radio resource allocation in the hierarchical fog-computing C-RAN network for enhancing the latency performance, which has unfortunately not been addressed in existing works. 
	
	In this paper, we investigate the computational task offloading scheme in a hierarchical fog-computing C-RAN network, which consists of three tiers of computational services: MEC server (MEC-L) in RUs, MEC server (MEC-H) in DUs, and the cloud computing in CUs. 
	Our target is to minimize the total delay of the computation tasks by optimizing the receive beamforming vectors, task allocation, computing speed for offloaded tasks in each server and the transmission bandwidth split of fronthaul links.
	The formulated mixed integer non-linear problem (MINLP) problem is transformed into a convex problem by relaxing the integer constraints, and it is then solved by applying the Lagrange dual method.
	Simulation results are provided to illustrate the delay performance improvement and the impacts of the system parameters.

	\section{System Model}
	
	Consider the uplink of the fog-computing enabled C-RAN network consisting of a CU, $J$ DUs, $I$ RUs and $K$ users.
	Each user has a single transmit antenna and each RU is equipped with $M$ receiving antennas.
	Suppose that each user is served by its nearest RU, and the set of users that are served by RU $i$ is represented by $\mathcal{U}_i$.
	The RUs connected to DU $j$ is represented by $\mathcal{I}_j$.
	
\begin{figure}
		\centering
		\includegraphics[width=0.7\linewidth]{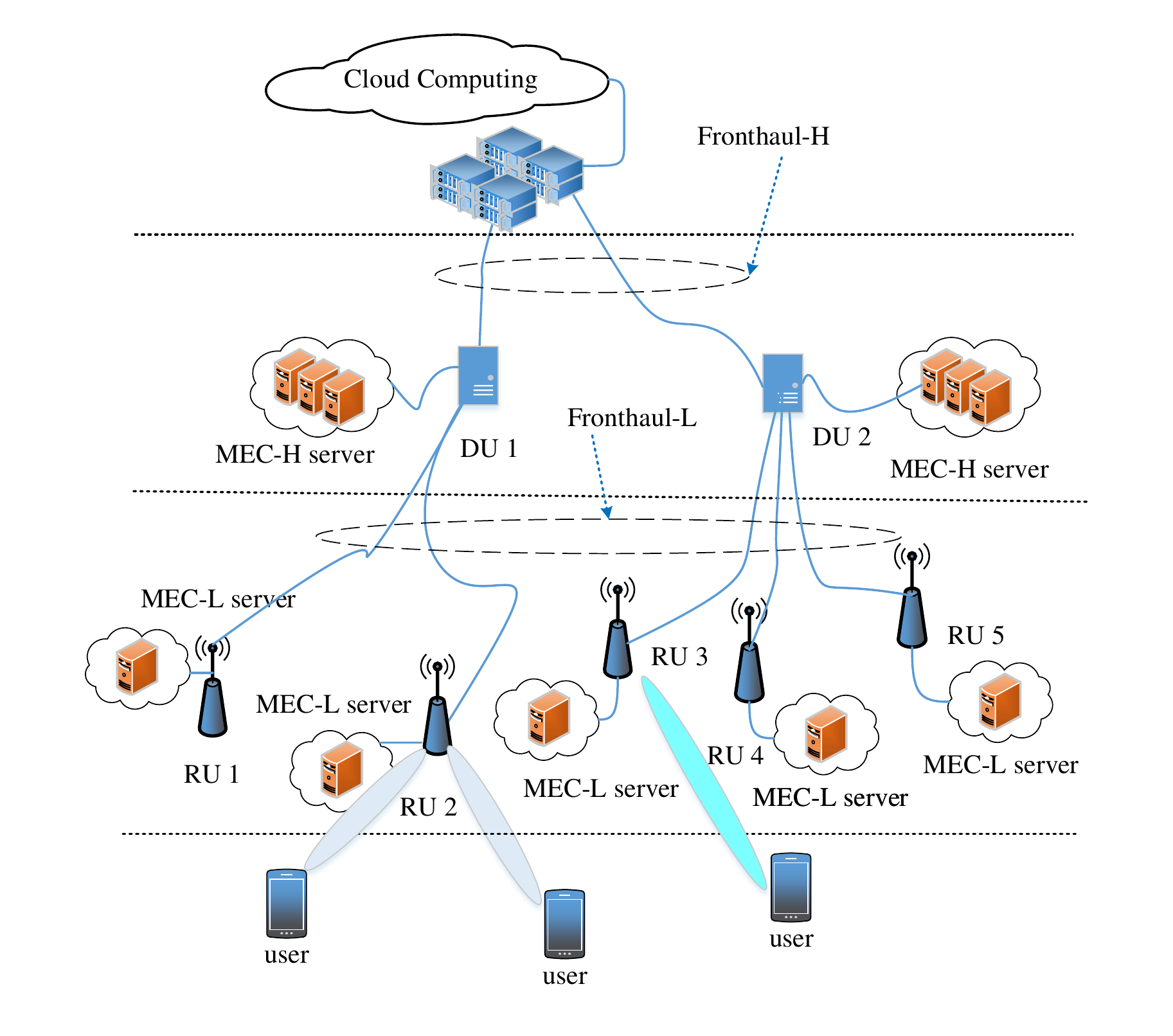}
		\vspace{-1em}
		\caption{An illustration of the C-RAN network equipped with multiple tiers of MEC services.}
		\vspace{-1em}
		\label{fig0}
	\end{figure}

	As shown in Fig. \ref{fig0}, multiple RUs can be connected with a single DU, and RUs are connected with DUs via bandwidth limited fronthaul links, which are denoted as fronthaul. 
	For distinction, the fronthaul links between DUs and the CU are denoted by midhaul.
	To provide the computation services, MEC servers with different computation capabilities are equipped in RUs and DUs, which are denoted by MEC-L and MEC-H, respectively.

	Suppose that user $k$ is served by RU $i$, i.e., $k \in \mathcal{U}_i$, and RU $i$ is connected to DU $j$, i.e., $i \in \mathcal{I}_j$.
	Suppose that orthogonal transmission spectrum bands are allocated to adjacent RUs for their the uplink transmissions.
	The wireless transmit channel from user $k$ to RU $i$ is denoted by ${\bm{h}}_{k,i} \in \mathbb{C}^{M \times 1}$.
	Then, the received signal vector at RU $i$ is given by 
	\begin{equation}
		{\bm y}_{i} = \sqrt{p_t}\sum_{k\in\mathcal{U}_i}{\bm{h}}_{k,i}s_{k,i} + {\bf{n}}_{i},  
	\end{equation}
	where $p_t$ is the transmit power of user $k$, $s_{k,i}$ is the transmit signal satisfying $\mathbb{E}[s_{k,i}^2]=1$ and the signals for different users are independent from each other and from the noise, i.e., $\mathbb{E}[{s_{k,i}s_{l,j}}] =0, \forall k \neq l, i \neq j$.
	
	Consider the linear receive beamforming strategy.
	Let ${\bm u}_{k,i}^{H}$ denote the receiving beamforming vector for user $k$ so that the estimated signal $\hat{s}_{k,i}$ is given by
	$\hat{s}_{k,i} = {\bm u}_{k,i}^{H} {\bm y}_{i}$.
	Then, the received SINR for user $k$ is 
	\begin{equation}
		\text{SINR}_{k,i}({\bm u}_{k,i}) = \frac{p_t|{\bm u}_{k,i}^{H}{\bm{h}}_{k,i}|^2 }{\sum_{l \neq k, l \in \mathcal{U}_i}p_t|{\bm u}_{k,i}^{H}{\bm{h}}_{l,i}|^2+|{\bm u}_{k,i}|^2\sigma^2},
	\end{equation}
	where $\sigma^2$ is the noise power of the Gaussian noise.

	Let $B_i$ denote the bandwidth assigned to RU $i$'s uplink.
	Then, the uplink achievable rate from user $k$ to RU $i$ is 
	\begin{equation}
		R_{k,i}({\bm u}_{k,i}) = B_i\log \left(1 + \text{SINR}_{k,i}({\bm u}_{k,i}) \right).
	\end{equation}

	Suppose that each user has a task to be executed, and the task of user $k$ is referred to as the task $k$.
	Let $D_k$ denote the data size of task $k$, and the required computing CPU frequency cycles are proportional to the task data size by the coefficient $\zeta_k$, i.e., the required computing cycle of task $k$ is $\zeta_kD_k$.


	As shown in Fig. \ref{fig0}, the places that task $k$ can be offloaded to can be the following three places: the MEC-L server located in the RUs, the MEC-H server located in the DUs and the cloud computing server connected to the CU.
	
	To represent the offloading decision in the proposed fog-computing network, the binary variables $x_k^{L}$,$x_k^{H}$ and $x_k^{C}$ are adopted to indicate the MEC-L offloading, MEC-H offloading and the cloud offloading, respectively, i.e.,
	\begin{equation}
		C1: x_k^{L},x_k^{H},x_k^{C} \in \{0,1\}.
	\end{equation}

	Then, the delay for each offloading decision is analyzed as follows.
	
	1) MEC-L offloading: If task $k$ is offloaded to the MEC-L located in RU $i$, the transmission delay is given by
	$T^t_{k} = \frac{D_k}{R_{k,i}} \label{delay1}$.
	Letting $f_{k,i}^{L}$ denote the computing speed of the MEC-L server located in RU $i$ for task $k$, the computing delay is given by
	$T^{C,L}_{k,i} = \frac{\zeta_kD_k}{f_{k,i}^{L}}$.
	Then, the total delay of the MEC-L offloading is $T^t_{k}+ T^{C,L}_{k,i}$.

	2) MEC-H offloading: 
	The fronthaul link which connects the RU $i$ is denoted by fronthaul $i$.
	The transmission bandwidth for task $k$ in fronthaul $i$ is denoted by $B^L_{k,i}$, and the spectrum efficiency of fronthaul $i$ is denoted as $R_i^L$, which is pre-determined according to the specific requirements of the fronthaul links.
	Then, the transmission delay in the fronthaul $i$ for task $k$ is given by
	$T^{L}_{k,i} = \frac{D_k}{B^L_{k,i}R^{L}_{i}}$.
	Letting $f_{k,j}^{H}$ denote the computing speed of the MEC-H server located in DU $j$, the computing delay is given by
	$T^{C,H}_{k,j} = \frac{\zeta_kD_k}{f_{k,j}^{H}}$.
	Then, the total delay for task $k$ adopting the MEC-H offloading is $T^t_{k}+ T^{L}_{k,i}+ T^{C,H}_{k,j}$.

	3) Cloud offloading: If task $k$ is offloaded to the cloud computing service located in CUs, the computing delay in the cloud server can be calculated as 
	$T_{k}^C = \frac{\zeta_kD_k}{f_k^C}$,
	where $f_k^C$ is the computing frequency provided by the cloud server to executing task $k$.
	It is assumed that the transmission bandwidth for task $k$ in midhaul $j$ is represented by $B_{k,j}^H$, and the spectrum efficiency of midhaul $j$ is denoted as $R^{H}_{j}$.
	The transmission delay in the midhaul $j$ is given by
	$T^{H}_{k,j} = \frac{D_k}{B_{k,j}^H R^{H}_{j}}$.
	To access the cloud computing located in CUs, the task $k$ should experience the transmit delays caused by both the fronthaul $i$ and midhaul $j$.
	Then, the total delay of cloud offloading for task $k$ is 
	$T^t_{k} + T^{L}_{k,i}+T^{H}_{k,j}+T_{k}^C$.

	\section{Problem Formulation}
	Obviously, each task may only be conducted by one place for computing, so that 
	\begin{equation}
		C2: x_k^{L} + x_k^{H} + x_k^{C} = 1.
	\end{equation}
	
	As the computation capabilities of the MEC-L, MEC-H and the cloud computing are different, the computing speeds have the following constraints:
	\begin{align}
		&C3: \sum_{k \in \mathcal{U}_i}x_k^{L} f_{k,i}^{L} \leq F^{L}_{i},\\
		&C4: \sum_{i \in \mathcal{I}_j}\sum_{k \in \mathcal{U}_i} x_k^{H} f_{k,j}^{H} \leq F^{H}_{j},\\
		&C5: \sum_{j}\sum_{i \in \mathcal{I}_j}\sum_{k \in \mathcal{U}_i} x_k^{C} f_{k}^{C} \leq F^{C}.
	\end{align}
	
	As the midhaul and fronthaul are bandwidth limited, the following constraints are introduced:
	\begin{align}
		&C6: \sum_{k \in \mathcal{U}_i}(x_k^{H} + x_k^{C})B^L_{k,i} \leq B^{L}_{i},\\
		&C7: \sum_{i \in \mathcal{I}_j} \sum_{k \in \mathcal{U}_i} x_k^{C} B_{k,j}^H\leq B^{H}_{j}.
	\end{align}
	
	Then, the optimization problem can be formulated as 
	\begin{subequations}\label{pro1}
		\begin{align}
			\underset{ \underset{\{x_k^{L},x_k^{H},x_k^{C}\},\{\bm u_{k}\},}{\{f_{k,i}^{L},f_{k,j}^{H},f_{k}^{C}\},\{B_{k,i}^{L},B_{k,j}^{H}\} }}{\text{min}}  
			&\quad \sum_{j}\sum_{i \in \mathcal{I}_j} \sum_{k \in \mathcal{U}_i}T_k    \label{obj1}  \\
			\text{s.t.} & \quad C1-C7.
		\end{align}
	\end{subequations}
	
	Clearly, Problem (\ref{pro1}) is a MINLP, which is nonconvex in general and not easy to solve directly. 
	However, it can be inferred that the formulated problem can be divided into two sub-problems, as the optimization variable $\{{\bm u}_{k,i}\}$ is not coupled with other optimization variables.
	
	\section{Solution Analysis}

	\subsection{Computing frequency,  bandwidth and offloading decision}
	
	The computing frequency, transmission bandwidth and the offloading decision can be obtained by solving the following problem:
	\begin{subequations}\label{pro2}
		\begin{align}
			\underset{\underset{\{x_k^{L},x_k^{H},x_k^{C}\}}{\{f_{k,i}^{L},f_{k,j}^{H},f_{k}^{C}\}, \{B_{k,i}^L, B_{k,j}^H\}}}{\text{min}}  
			& \sum_{j}\sum_{i \in \mathcal{I}_j} \sum_{k \in \mathcal{U}_i} \tilde{T}_k  \label{obj2}  \\
			\text{s.t.} & \quad C1-C6,
		\end{align}
	\end{subequations}
	where $\tilde{T}_k$ is given by
	\begin{equation}
	\tilde{T}_k = x_{k}^L\frac{\zeta_kD_k}{f_{k,i}^{L}} 
	+ x_{k}^H \frac{\zeta_kD_k}{f_{k,j}^{H}}
	+ x_{k}^C \frac{\zeta_kD_k}{f_{k}^{C}}
	+ x_{k}^H \frac{D_k}{B_{k,i}R^{L}_{i}}
	+ x_{k}^C \left(\frac{D_k}{B_{k,i}R^{L}_{i}}+ \frac{D_k}{B_{k,j}R^{H}_{j}}\right).
	\end{equation}

	It can be seen that Problem (\ref{pro2}) is still non-convex and hard to solve.
	In the following, we first relax the integer constraints $C1$ and introduce the following variable substitutions:
	\begin{align}
		& x_k^L f_{k,i}^L = a_{k,i}, x_k^H f_{k,j}^H = b_{k,j}, x_k^C f_{k}^C = c_{k},\\
		& x_k^H B_{k,i}^L = \alpha_{k,i}, x_k^C B_{k,j}^H = \beta_{k,j},x_k^C B_{k,i}^H = \gamma_{k,i}.
	\end{align}
	
	Note that $f(x)= \frac{a}{x}$ is convex with respect to $x$, so that its perspective function $g(t,x)= tf(\frac{t}{x})$ is convex with respect to $(t,x)$. 
	Then, by using  the defined substitutions, we can obtained the following convex problem ()\ref{pro3}), and the Lagrange dual method can be applied.
	
	$\tilde{T}_k$ is transformed into
	\begin{equation}
	\tilde{T}_k= (x_{k}^L)^2 \frac{\zeta_kD_k}{a_{k,i}} 
	+ (x_{k}^H)^2\frac{\zeta_kD_k}{b_{k,i}}
	+ (x_{k}^C)^2\frac{\zeta_kD_k}{c_{k}}
	+ (x_{k}^H)^2 \frac{D_k}{\alpha_{i,k}R^{L}_{i}}
	+ (x_{k}^C)^2 \left(\frac{D_k}{\gamma_{i,k}R^{L}_{i}}+ \frac{D_k}{\beta_{i,j}R^{H}_{j}}\right).
	\end{equation}
	
	Hence, Problem (\ref{pro2}) is transformed into
	\begin{subequations}\label{pro3}
		\begin{align}
			\!\!\!\!\!\!\!\!\!\!\underset{\underset{\{x_k^{L},x_k^{H},x_k^{C}\}}{\{a_{k,i},b_{k,j},c_{k}\}, \{\alpha_{k,i},\beta_{k,j},\gamma_{k,i}\}}}{\text{min}}  
			&\quad 
			\sum_{j}\sum_{i \in \mathcal{I}_j} \sum_{k \in \mathcal{U}_i} \tilde{T}_k \label{obj3}  \\
			\text{s.t.} \quad & x_k^{L} + x_k^{H} + x_k^{C} = 1, \forall k \in \mathcal{U}_i, \label{St_inger}\\
			& \sum_{k \in \mathcal{U}_i} a_{k,i} \leq F^{L}_{i},\forall k \in \mathcal{U}_i,  \label{st1}\\
			& \sum_{i \in \mathcal{I}_j}\sum_{k \in \mathcal{U}_i} b_{k,j} \leq F^{H}_{j},\forall i \in \mathcal{I}_j,\\
			& \sum_{j}\sum_{i \in \mathcal{I}_j}\sum_{k \in \mathcal{U}_i} c_k\leq F^{C},\\
			& \sum_{k \in \mathcal{U}_i}(\alpha_{k,i}+ \gamma_{k,i})\leq B^{L}_{i},\\
			& \sum_{i \in \mathcal{I}_j} \sum_{k \in \mathcal{U}_i} \beta_{k,j} \leq B^{H}_{j},\label{st2}\\
			& x_k^{L},x_k^{H},x_k^{C} \in [0,1].
		\end{align}
	\end{subequations}

	We introduce the non-negative variables $\mu_i,\lambda_j,\rho, \nu_i, \xi_j$ to indicate the constraints (\ref{st1})-(\ref{st2}), respectively. Then, the Lagrange function can be obtained.

	To obtain the optimal values of $({a_{k,i}}^*,{b_{k,i}}^*,{c_{k}}^*)$, we take the first order derivatives of the Lagrange function with respect to $a_{k,i}$, $b_{k,i}$, and $c_{k}$, respectively. 
	Then, we can obtain the following equations:
	\begin{align}
		&\frac{\partial \mathcal{L}}{\partial a_{k,i}} 
		= - (x_k^L)^2\frac{\zeta_kD_k}{a^2_{k,i}} + \mu_i, \label{Ldiv1} \\
		&\frac{\partial \mathcal{L}}{\partial b_{k,j}} 
		= - (x_k^H)^2\frac{\zeta_kD_k}{b^2_{k,j}} + \lambda_j,\label{Ldiv2}\\
		&\frac{\partial \mathcal{L}}{\partial c_{k}} 
		= - (x_k^H)^2\frac{\zeta_kD_k}{c^2_{k}} + \rho,\label{Ldiv3}
	\end{align}
	
	It is observed that if ${x_{k}^L}^*=0$, then we have ${a_{k,i}}^*=0$, if ${x_{k}^H}^*=0$, then ${b_{k,i}}^*=0$, and if ${x_{k}^C}^*=0$, then ${c_{k}}^*=0$. 
	Then, if $x_{k}^L,x_{k}^H,x_{k}^C \neq 0$, according to (\ref{Ldiv1}) -(\ref{Ldiv3}), it is inferred that 
	\begin{equation}
		{a^*_{k,i}} = {x_{k}^L}^*\sqrt{\frac{\zeta_kD_k}{\mu_i}},
		{b^*_{k,j}} = {x_{k}^H}^*\sqrt{\frac{\zeta_kD_k}{\lambda_j}},
		{c^*_{k}} = {x_{k}^C}^*\sqrt{\frac{\zeta_kD_k}{\rho}}.  
	\end{equation}
	
	To obtain the optimal values of $({\alpha_{k,i}}^*,{\beta_{k,i}}^*,{\gamma_{k}}^*)$, the first order derivatives of the Lagrange function with respect to $\alpha_{k,i}$, $\beta_{k,i}$, and $\gamma_{k}$ are given by
	\begin{align}
		&\frac{\partial \mathcal{L}}{\partial \alpha_{k,i}} 
		= - (x_k^H)^2\frac{D_k}{{\alpha_{k,i}}^2 R_i^L} + \nu_i, \label{Ldiv7} \\
		&\frac{\partial \mathcal{L}}{\partial \beta_{k,j}} 
		= - (x_k^C)^2\frac{D_k}{{\beta_{k,j}}^2 R_j^H} + \xi_j,\label{Ldiv8}\\
		&\frac{\partial \mathcal{L}}{\partial \gamma_{k,i}} 
		= - (x_k^C)^2\frac{D_k}{(\gamma_{k,i})^2 R_i^L} + \nu_i.\label{Ldiv9}
	\end{align}
	
	Similarly, we can infer that if ${x_{k}^H}^*=0$, then we have ${\alpha_{k,i}}^*=0$ and ${\gamma_{k,i}}^*=0$, and if ${x_{k}^C}^*=0$, then ${\beta_{k,j}}^*=0$.
	Otherwise, according to (\ref{Ldiv7})-(\ref{Ldiv9}), one obtains
	\begin{equation}
		{\alpha^*_{k,i}} = {x_{k}^H}^*\sqrt{\frac{D_k}{R_i^L\nu_i}},
		{\beta^*_{k,j}} = {x_{k}^C}^*\sqrt{\frac{D_k}{R_j^H\xi_j}},
		{\gamma^*_{k,i}} = {x_{k}^C}^*\sqrt{\frac{D_k}{R_i^L\nu_i}}.  
	\end{equation}
	
	Then, we need to determine the optimal offloading decision $({x_{k}^L}^*,{x_{k}^H}^*,{x_{k}^C}^*)$.
	By taking the first order derivatives of the Lagrange $\mathcal{L}$ with respect to ${x_{k}^L},{x_{k}^H},{x_{k}^C}$ respectively, we have the following equations:
	\begin{align}
		&\frac{\partial \mathcal{L}}{\partial x_k^L} 
		= 2\frac{\zeta_kD_k}{\frac{a_{k,i}}{x_k^L}},  \label{Ldiv4}\\
		&\frac{\partial \mathcal{L}}{\partial x_k^H} 
		= 2\frac{\zeta_kD_k}{\frac{b_{k,j}}{x_k^H}} + 2\frac{D_k}{\frac{\alpha_{k,i}}{x_k^H} R_i^L}, 
		\label{Ldiv5}  \\
		&\frac{\partial \mathcal{L}}{\partial x_k^C}
		= 2\frac{\zeta_kD_k}{\frac{c_{k}}{x_k^C}} + 2\frac{D_k}{\frac{\gamma_{k,i}}{x_k^C} R_i^L}+ 2\frac{D_k}{\frac{\beta_{k,j}}{x_k^C} R_j^H}. \label{Ldiv6}
	\end{align}
	
	For simplicity, define the following denotations:
	\begin{align}
		&L_{k} = \frac{\partial \mathcal{L}}{\partial x_k^L}\left(\frac{a_{k,i}^*}{{x_k^L}^*}\right), H_k = \frac{\partial \mathcal{L}}{\partial x_k^H}\left(\frac{b_{k,i}^*}{{x_k^H}^*},\frac{{\alpha_{k,i}}^*}{{x_k^H}^*}\right),\nonumber\\
		&C_k = \frac{\partial \mathcal{L}}{\partial x_k^C}\left(\frac{c_{k}^*}{{x_k^C}^*},\frac{{\beta_{k,j}}^*}{{x_k^C}^*},\frac{{\gamma_{k,i}}^*}{{x_k^C}^*}\right).
	\end{align}
	
	Consequently, according to constraint (\ref{St_inger}), the solution $({x_{k}^L}^*,{x_{k}^H}^*,{x_{k}^C}^*)$ can be inferred as 
	\begin{equation}
		\left\{\begin{array}{llll}
			{x_{k}^L}^*=1,{x_{k}^H}^*=0,{x_{k}^C}^*=0, \text{if } L_k < \min\left\{H_k, C_k\right\}, \\
			{x_{k}^L}^*=0,{x_{k}^H}^*=1,{x_{k}^C}^*=0, \text{if } H_k < \min\left\{L_k, C_k\right\}, \\
			{x_{k}^L}^*=0,{x_{k}^H}^*=0,{x_{k}^C}^*=1, \text{if } C_k< \min\left\{L_k, H_k\right\}. 
		\end{array}\right.
	\end{equation}

	Note that the values of the dual variables $\mu_i$, $\lambda_j$, $\rho$, $\nu_i$ and $\xi_j$ can be determined by the sub-gradient method.
	According to \cite[Proposition 6.3.1]{bertsekas1999nonlinear}, the sub-gradient method converges to the optimal solution to Problem (\ref{pro3}) for sufficiently small step-sizes.
	
	\subsection{Receiving beamforming vector} 
	
	The receiving beamforming vector can be determined by the following problem.
	\begin{equation}\label{pro4}
		\underset{\{\bm u_{k,i}\}}{\text{min}}  
		\quad \sum_{j}\sum_{i \in \mathcal{I}_j} \sum_{k \in \mathcal{U}_i} \frac{D_k}{R_{k,i}({\bm u}_{k,i})}.    
	\end{equation}
	
	Then, it is readily to see that the optimal solution $\{\bm u_{k,i}\}$ to Problem (\ref{pro5}) is equivalent to 
	\begin{equation}\label{pro5}
		\underset{\bm u_{k,i}}{\text{max}}  
		\quad \text{SINR}_{k,i}({\bm u}_{k,i}).
	\end{equation} 
	
	Then, as it has been clearly shown in \cite{Palomar.2003} that maximizing the SINR is clearly equivalent to minimizing the MSE.
	It is readily obtained that the optimal receive beamforming vector is 
	\begin{equation}
		{\bm u}_{k,i} = \left(\sigma^2\bm{I}_{M} + \sum_{l \in \mathcal{U}_i} \bm{h}_{l,i}\bm{h}_{l,i}^H \right)^{-1}\bm{h}_{k,i},
	\end{equation}
	where $\bm{I}_{M}$ is the identity matrix with size $M \times M$.

	\section{Simulation Results}
	
	In our simulations, the $K$ users are randomly deployed in the fog-computing network.
	The wireless transmission channel is modelled by the 3GPP Spatial channel model (SCM) \cite{SCMmodel} for MIMO simulations. 
	For the computation tasks, the size of each task is uniformly generated in the range $[5,30]$ Mbits, and the required computation frequency coefficient is uniformly distributed in [$0.1$,$10$].
	Most of the simulation parameters are presented in Table \ref{tab1}.
	All the presented simulation results are obtained by averaging over $1000$ random realizations.
	
	\begin{table}
		\centering
		\caption{The simulation parameters}
		\label{tab1}
		\begin{tabular}{|l|l|l|}
			\hline
			Parameters	& Value   \\
			\hline
			Uplink Bandwidth $B_i$ & 10 MHz \\
			Noise power density	& $-174$ dBm/Hz   \\
			Transmit power of users $p_t$ & $35$ dBm  \\ 
			Convergence precision $\epsilon$ & $10^{-4}$ \\
			Spectrum efficiency $R_i^L$ and $R_j^H$ & $3$ bits/Hz \\
			Number of receiving antennas $M$ & 10\\
			Number of DUs  & 4 \\
			Number of RUs & 10 \\
			Cloud computing frequency $F^C$ & $5 \times 10^3$ GHz \\
			\hline
		\end{tabular}
	\end{table} 
	
	The proposed fog-computing network structure is  labeled as ``Fog''.
	For performance comparison, we consider the following cases: 1) Without the MEC-L and MEC-H servers, this network is only equipped with the cloud computing service, which is referred to the ``Cloud'' scheme; 2) By removing the MEC-L severs in RUs, the tasks can only be offloaded into the MEC-H servers and the cloud, and this structure is denoted by the ``Cloud+DU'' scheme; 3)There is no MEC-H servers in DUs, which is labeled by the ``Cloud+RU'' scheme.

	%
	%

	\begin{figure}
		\centering
		\vspace{-0.5em}
		\includegraphics[width=0.7\linewidth]{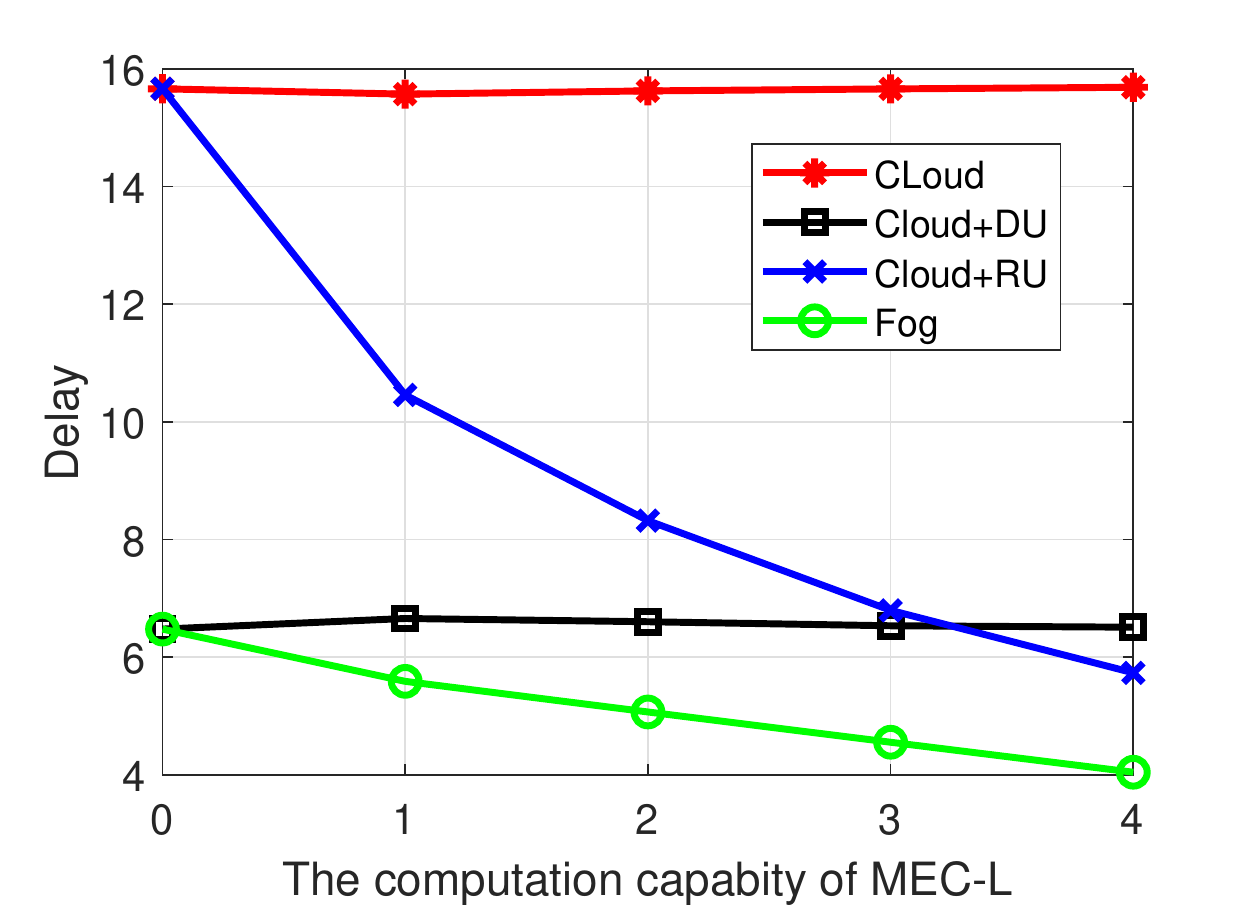}
		\caption{The impact of MEC-L servers' computation capabilities on the delay performance with $F_j^H = 25$ GHz,$B^{L}_{i} =300$MHz and $B^{H}_{j} =500$MHz.}
		\label{figs2}
	\end{figure}
	
	\begin{figure}
		\centering
		\vspace{-0.5em}
		\includegraphics[width=0.7\linewidth]{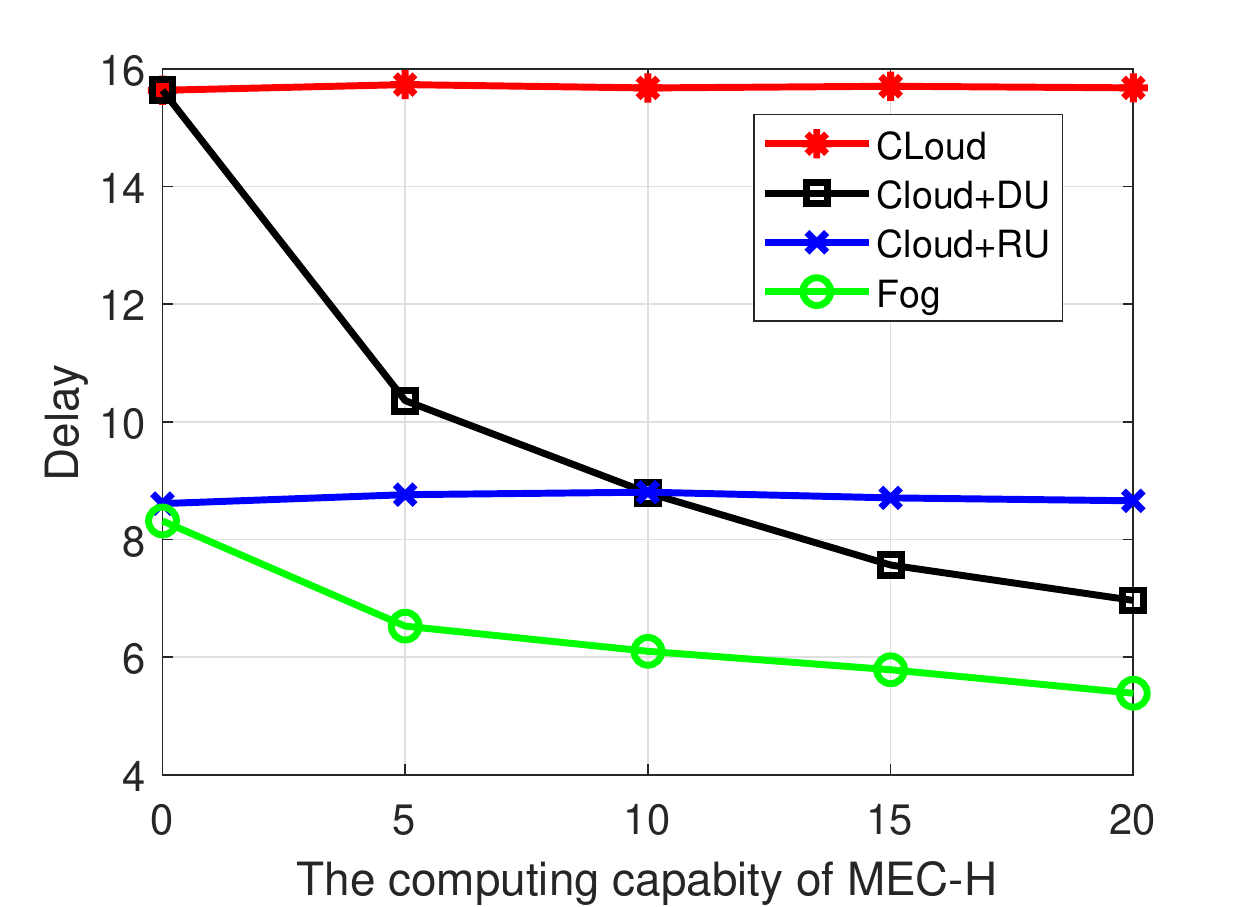}
		\caption{The impact of MEC-H servers' computation capabilities on the delay performance with $F_i^L =2$ GHz, $B^{L}_{i} =300$MHz and $B^{H}_{j} =500$MHz.}
		\label{figs3}
	\end{figure}
	
	Fig. \ref{figs2} and Fig. \ref{figs3} show how the delay performance behaves when the computation capabilities of the MEC-L servers and MEC-H servers change, respectively.
	It is readily to see that the proposed fog-computing scheme always achieves the best delay performance among all schemes. 
	In Fig. \ref{figs2}, the total delay of the ``Cloud+RU'' and the ``Fog'' decreases with the computation capabilities of the MEC-L.
	Meanwhile, in Fig. \ref{figs3}, the total delay of the ``Cloud+DU'' and the ``Fog'' decreases with the computation capabilities of the MEC-H.
	Consequently, it is inferred that the relative size of the computation capabilities of the MEC-L and MEC-H servers has a great influence on the delay performance achieved by the ``Cloud+RU'' and the ``Cloud+DU'' schemes.

	\begin{figure}
		\centering
		\vspace{-0.5em}
		\includegraphics[width=0.7\linewidth]{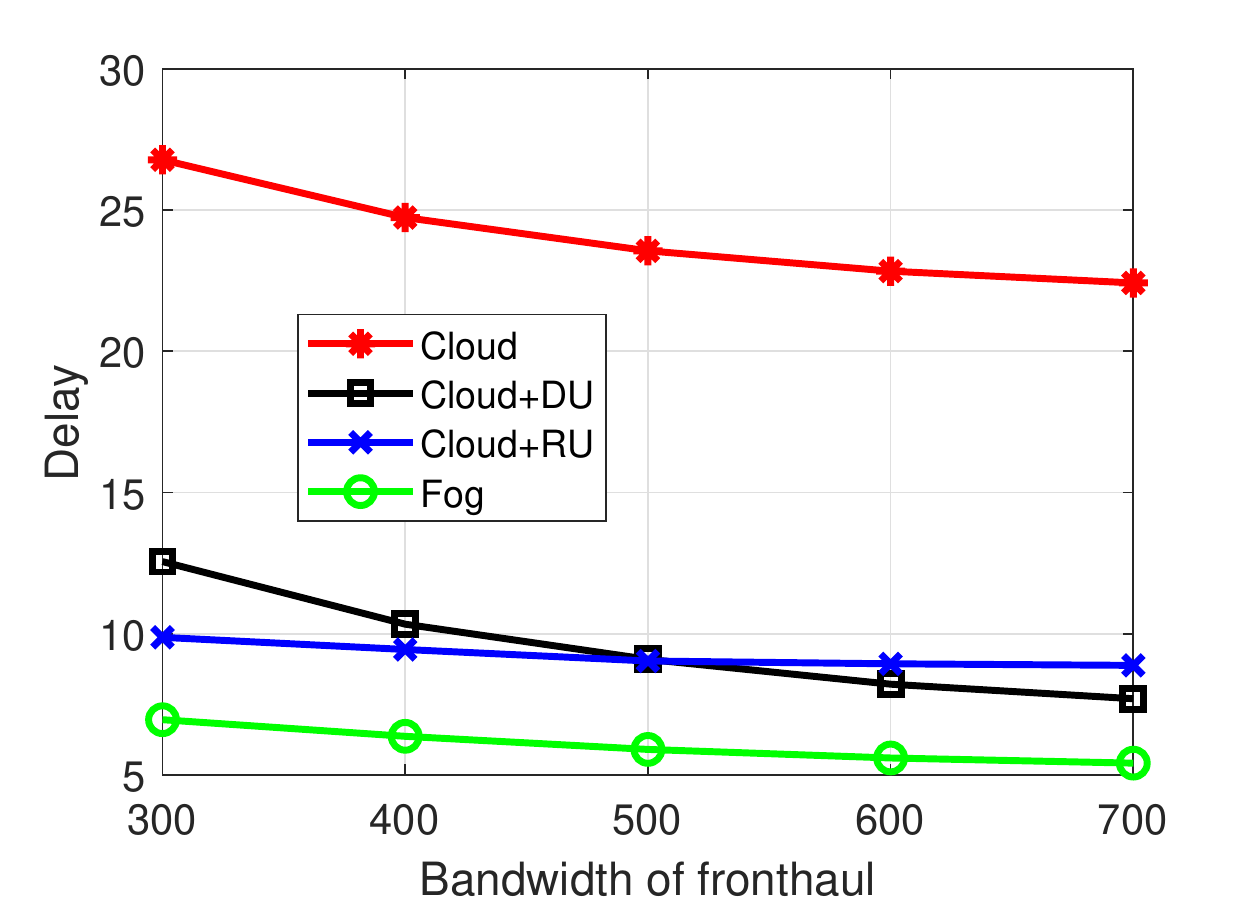}
		\caption{The impact of Midhaul bandwidth on the delay performance with $B^{H}_{j} =500$MHz.}
		\label{figs5}
	\end{figure}
	
	Fig. \ref{figs5} shows the delay performance versus the bandwidth of the fronthaul.
	In Fig. \ref{figs5}, the computation capabilities of MEC-L servers and MEC-H servers are randomly generated in the interval $[1,5]$ GHz and $[10,50]$ GHz.  
	It is observed in Fig. \ref{figs5} that the delay decreases with the fronthaul's bandwidth.
	Moreover, the impacts of fronthaul's bandwidth on the ``Cloud+DU'' scheme and the ``Cloud'' scheme are more significant.
	This is due to the fact that the task offloading in these two schemes experiences a longer transmission distance.
	Therefore, the ``Cloud+RU'' and ``Fog'' schemes have the delay performance superiority for the fronthaul's bandwidth limited network.
	
	\begin{figure}
		\centering
		\vspace{-0.5em}
		\includegraphics[width=0.7\linewidth]{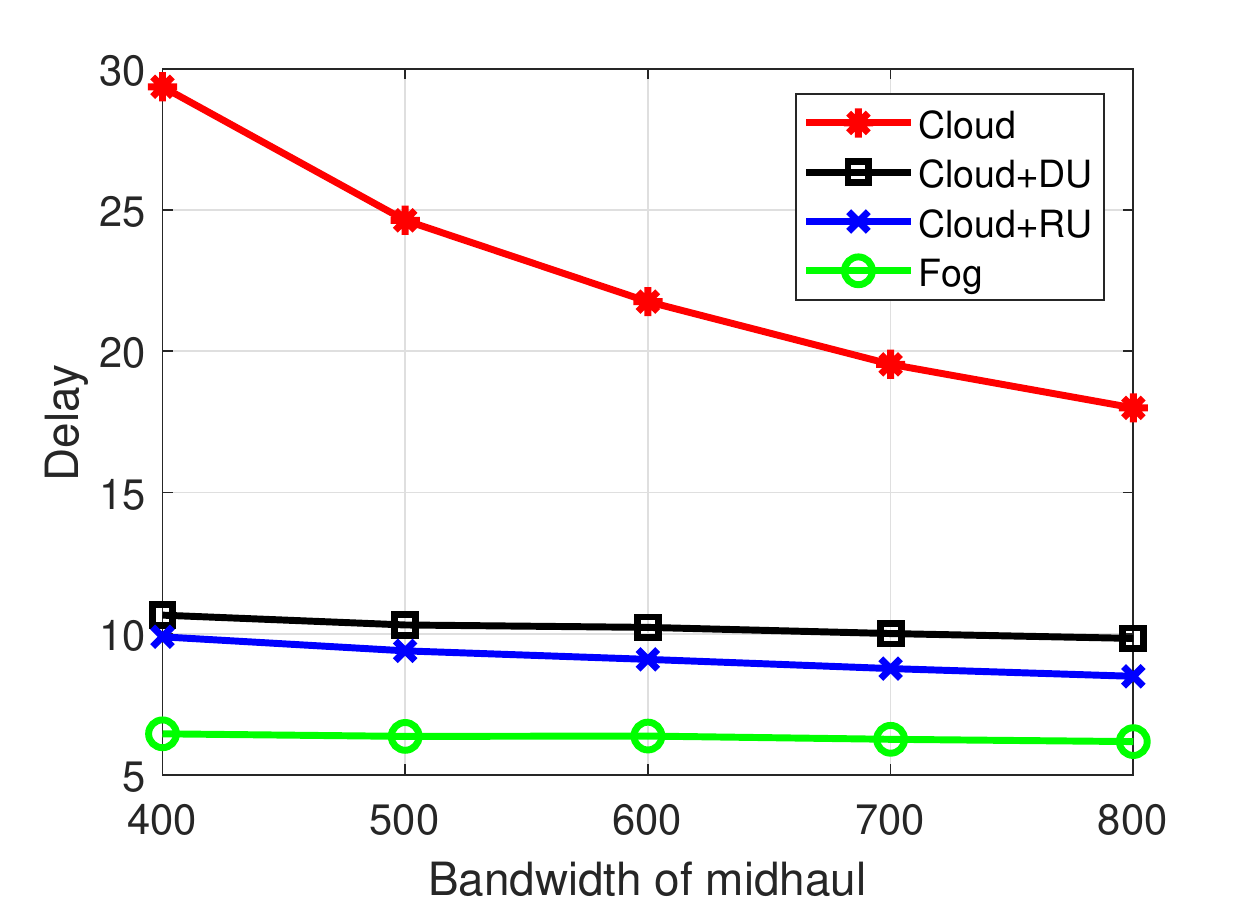}
		\caption{The impact of fronthaul bandwidth on the delay performance with $B^{L}_{i} =400$MHz.}
		\label{figs6}
	\end{figure}
	
	Fig.\ref{figs6} shows how the delay performance changes as the bandwidth of the midhaul increases.
	It is observed that the performance of the ``Cloud'' scheme is significantly influenced by the midhaul's bandwidth.
	This is because the tasks offloaded to the cloud need to be transmitted through the midhaul.
	Meanwhile, the delay performance of the other three schemes slightly decreases with the bandwidth of midhaul. 
	This implies that most of the tasks are executed by the MEC-L servers and the MEC-H servers in the ``Cloud+DU'', ``Cloud+RU'' and ``Fog'' schemes, which makes these schemes attractive for the midhaul's bandwidth limited networks.

	%
	
	\section{Conclusion}
	
	In this work, we have investigated the delay performance of the task offloading in a hierarchical fog-computing C-RAN network, where the computational tasks can be offloaded to three tiers: MEC-L in RUs, MEC-H in DUs, and the cloud computing in CUs.
	It is shown that the hierarchical fog-computing C-RAN network can significantly improve the delay performance of task offloading, in comparison to the two-tier fog computing scheme and the cloud computing scheme. 
	Furthermore, it is interesting to see that placing all the computational resources in RUs is not always the most efficient way. 
	When the bandwidth of the fronthaul link is large enough and the total budget of the computation resource is very limited, it is better to gather the computational resources in DUs instead of distributing them into RUs.
	
	\section*{Acknowledgment}
	This work was supported in part by the National Natural Science Foundation of China under Grants No. 61871128, No. 61971129, No. 61960206005,No.61801227, Basic Research Project of Jiangsu Provincial Department of Science and Technology under Grant No. BK20190339, natural science foundation for colleges and universities in Jiangsu Province under Grant No. 18KJB510022, and the UK Royal Society Newton International Fellowship under Grant NIF$ \verb|\|$ R1$ \verb|\|$180777.

	\bibliographystyle{ieeetran}
	\bibliography{Refer_0529}

\end{document}